\documentclass[twocolumn,showpacs,prb]{revtex4}
\usepackage{epsfig}
\usepackage{graphicx}
\newcommand{\vnabla}{{\mbox{\boldmath$\nabla$}}}

\newcommand{\vR}{{\mbox{\boldmath$R$}}}
\newcommand{\vJ}{{\mbox{\boldmath$J$}}}
\newcommand{\vD}{{\mbox{\boldmath$D$}}}
\newcommand{\vn}{{\mbox{\boldmath$n$}}}

\newcommand{\vk}{{\mbox{\boldmath$k$}}}

\newcommand{\vm}{{\mbox{\boldmath$m$}}}

\newcommand{\vsk}{{\small \mbox{\boldmath$k$}}}

\newcommand{\vg}{\mbox{\boldmath$g$}}
\newcommand{\vq}{\mbox{\boldmath$q$}}

\newcommand{\hvg}{\hat{\mbox{\boldmath$g$}}}

\newcommand{\vsig}{\mbox{\boldmath$\sigma$}}

\newcommand{\mhx}{\hat{\mbox{\boldmath$x$}}}

\newcommand{\vH}{\mbox{\boldmath$H$}}
\newcommand{\vB}{\mbox{\boldmath$B$}}
\newcommand{\vA}{\mbox{\boldmath$A$}}

\begin{document}

\title{Helical vortex phase in the non-centrosymmetric CePt$_3$Si}
\author{R.P. Kaur, D.F. Agterberg, and M. Sigrist}
\address{Department of Physics, University of Wisconsin-Milwaukee, Milwaukee, WI 53211}
\address{Theoretische Physik ETH-H¨onggerberg CH-8093 Z¨urich, Switzerland}

\begin{abstract}
We consider the role of magnetic fields on the broken inversion
superconductor CePt$_3$Si. We show that upper critical field for a
field along the c-axis exhibits a much weaker paramagnetic effect
than for  a field applied perpendicular to the c-axis. The
in-plane paramagnetic effect is strongly reduced by the appearance
of helical structure in the order parameter. We find that to get
good agreement between theory and recent experimental measurements
of $H_{c2}$, this helical structure is required.  We propose a
Josephson junction experiment that can be used to detect this
helical order. In particular, we predict that Josephson current
will exhibit a magnetic interference pattern for a magnetic field
applied {\it perpendicular} to the junction normal. We also
discuss unusual magnetic effects associated with the helical
order.

\end{abstract} \maketitle

The recently discovered heavy fermion superconductor CePt$_3$Si
\cite{bau04} has triggered many experimental and theoretical
studies\cite{fri04,sam04,sax04,ser04,min04,sam04-2,yas04,yog04}.
There are two features which have caused this attention: the
absence of inversion symmetry; and the comparatively high upper
critical magnetic field ($H_{c2}$). Broken inversion symmetry
(parity) has a pronounced effect on the quasiparticle states
through the splitting of the two spin degenerate bands. This
influences the superconducting phase, which usually relies on the
formation of pairs of electrons in degenerate quasiparticle states
with opposite momentum. The availability of such quasiparticle
states is usually  guaranteed by time reversal and inversion
symmetries (parity) \cite{and59,and84}. It is relatively easy to
remove time reversal symmetry, {\it e.g.} by a magnetic field, and
the physical consequences of this have been well studied. However,
parity is not so straightforwardly manipulated by external fields.
Superconductivity in materials without inversion center therefore
provides a unique opportunity in this respect.

The large $H_{c2}\approx 4 T$ in CePt$_3$Si \cite{bau04,yas04}
implies that the Zeeman splitting must be non-negligible below $
T_c =0.75 K$ (the estimated paramagnetic limit is at $H_P\approx
1.2$ T). In a magnetic field, this superconductor has to form
Cooper pairs under rather odd circumstances. In particular, it is
no longer guaranteed that a state with momentum $\vk$ at the Fermi
surface has a degenerate partner at $ - \vk $. The state $\vk $
would rather pair with a degenerate state $ -\vk + \vq$ and in
this way generate an inhomogeneous superconducting phase. We argue
below that recent $H_{c2}$ measurements \cite{yas04} suggest that
this is the  case in CePt$_3$Si. These measurements show that,
while the upper critical field is basically isotropic close to $
T_c$, a small anisotropy appears at lower temperature \cite{yas04}
($H_{c2}^c/H_{c2}^{ab}=1.18$ at $T=0$). The apparent absence of a
paramagnetic limit in CePt$_3$Si can be explained by lack of
inversion symmetry even if the pairing has $s$-wave symmetry
\cite{bul76,fri04,gor01}. However, these works indicate that
suppression of paramagnetism is very anisotropic and the
application of this theory to CePt$_3$Si would indicate no
paramagnetic suppression for the field along the $c$-axis, but a
suppression for the field in-plane ($H_P^{ab}\approx 1.7 T$). The
relative lack of anisotropy in the experimental data is surprising
in this context.

In this letter we examine the mixed phase close to the upper
critical field. Using the crystal symmetry of CePt$_3$Si, we show
that the high-field superconducting phase has pronounced
differences for field-directions parallel and perpendicular to the
c-axis. For the field parallel to the c-axis, the paramagnetic
limiting is basically absent and the vortex phase is quite
conventional. While for the perpendicular field direction, the
field can induce a phase which gives rise to an additional phase
factor in the many body wavefunction $ e^{i \vq \cdot \vR} $ with
$ \vq $ perpendicular to the applied field: a helical vortex
phase.
We also propose a Josephson junction experiment that can be used
to detect this helical phase factor and discuss a transverse
magnetization related to the helical phase.

We use the single particle Hamiltonian
\begin{equation}
{\cal H}_0 = \sum_{\vsk,s,s'} \left[ \xi_{\vsk} \sigma_0 + \alpha
  \vg_{\vsk} \cdot \vsig +\mu_B\vH \cdot \vsig \right]_{ss'} c_{\vsk s}^{\dag} c_{\vsk s'}
\label{eq-1}
\end{equation}
where $ c_{\vsk s}^{\dag} $ ($ c_{\vsk s} $) creates (annihilates)
an electron with  momentum $ \vk $  and spin $s$. The band energy
$ \xi_{\vsk} = \epsilon_{\vsk} - \mu $ is measured relative to the
chemical potential $ \mu $, $ \alpha \vg_{\vsk} \cdot \vsig $
introduces the antisymmetric spin-orbit coupling with $ \alpha $
as a coupling constant (we set $ \langle \vg_{\vsk}^2 \rangle = 1
$ where $\langle ..\rangle $ is an  average over the Fermi
surface), and $\mu_B\vB \cdot \vsig$ gives the Zeeman coupling.
The crucial term in Eq.~\ref{eq-1} is $ \alpha \vg_{\vsk} \cdot
\vsig $, which is only permitted when inversion symmetry is broken
($\vg_{\vsk}$ satisfies $\vg_{\vsk}=-\vg_{-\vsk}$ due to time
reversal symmetry). This term destroys the usual two-fold spin
degeneracy of the bands by splitting the band into two
spin-dependent parts with energies $ E_{\vsk,\pm} = \xi_{\vsk} \pm
\alpha | \vg_{\vsk}| $. The spinors are determined by the
orientation of the corresponding $\vg_{\vsk}$. The general pairing
interaction is
\begin{equation} \begin{array}{l}
{\cal H}_{pair} = \displaystyle
\frac{1}{2} \sum_{\vsk, \vsk,\vq, s_i} V(\vk,\vk') \\
\quad \qquad \times \; c^{\dag}_{\vsk+\vq/2,s_1}
                     c^{\dag}_{-\vsk+\vq/2,s_2} c_{-\vsk'+\vq/2,s_2}
                     c_{\vsk'+\vq/2,s_1},
                     \end{array}
\end{equation}
expressed in the usual spin basis. We will work in the large
$\alpha$ limit so that the pairing problem becomes a real two-band
problem in the diagonal spinor ($\pm$) basis. To find the pairing
interaction in the $\pm$ basis, we diagonalize the single particle
Hamiltonian after which the two-band pairing interaction, for
$\vH=0$, is written in spinor form as
\begin{equation}
V= \frac{1}{2} V(\vk,\vk') \left( \begin{array}{cc} e^{i \phi_-} A_+ & e^{-i \phi_+} A_- \\
e^{i \phi_+} A_- & e^{-i \phi_-} A_+ \end{array} \right) \;
\label{twogap}
\end{equation}
where $ \phi_{\pm} = \phi_{\vk} \pm \phi_{\vk'} $ and $ A_{\pm} =
(1\pm \hvg_{\vsk}\cdot\hvg_{\vsk'})$ where we have taken
$\vg_{\vsk}=|\vg_{\vsk}|\big (\sin\theta_{\vsk}
\cos\phi_{\vsk},\sin\theta_{\vsk}\sin\phi_{\vsk},\cos\theta_{\vsk}\big
)$. Note that even for a
 spatially isotropic interaction, the two-band solution has both
a spin-triplet and a spin-singlet gap function when $\alpha\ne 0$
(this is a consequence of the broken parity symmetry
\cite{gor01}). We will consider the limit $\alpha>>\mu_B H$ and
keep only terms up to order $\mu_B H/\alpha$ (a good approximation
for CePt$_3$Si).  We restrict ourselves to choices of
$V(\vk,\vk')$ that corresponds to spin-singlet pairing in the
$\alpha=0$ limit. This restriction allows us to use
Eq.~\ref{twogap}, even if $\vH\ne0$, which considerably simplifies
the notation.

With the two band pairing interaction of Eq.~\ref{twogap}, the
linearized gap equation becomes
\begin{widetext}
\begin{equation}
\Delta_{\alpha}(\vk,\vq)=-T \sum_{n,{\bf
k'}}\sum_{\beta}V_{\alpha,\beta}(\vk,\vk')
G^0_{\beta}(\vk'+\vq/2,i \omega_n)G^0_{\beta}(-\vk'+\vq /2 ,-i
\omega_n)\Delta_{\beta}(\vk',\vq)
\end{equation}
\end{widetext}
where $ G_{\pm}^0(\vk,i\omega_n)= [
i\omega_n+\xi_{\vsk}\pm\alpha|\vg_{\vsk}|]^{-1} $. Setting
$\Delta_+(\vk,\vq)=e^{i\phi_{\vsk}}\tilde{\Delta}_+(\vk,\vq)$ and
$\Delta_-(\vk,\vq)= - e^{-i\phi_{\vsk}}\tilde{\Delta}_-(\vk,\vq)$
results in the simplified two-band equation with the interaction
\begin{equation}V=V(\vsk,\vsk')(\sigma_0 - \sigma_x)/2
 \label{twoband2}.
\end{equation}
The factors $\hvg$ in Eq.~\ref{twogap} do not appear because of
Pauli exclusion and the assumed singlet form of $V(\vsk,\vsk')$ in
the $\alpha=0$ limit. We denote the density of states on the two
Fermi surfaces by $N_+=N\cos^2 (\delta/2)$ and $N_-=N\sin^2
(\delta/2)$.
We can write
$\tilde{\Delta}_{\alpha}(\vk,\vq)=\psi_{\Gamma,\alpha}(\vq)f_{\Gamma}(\vk)$,
where Eq.~\ref{twoband2} implies that
$\psi_{\Gamma,+}(\vq)+\psi_{\Gamma,-}(\vq)=0$. With
$\psi_{\Gamma,+}(\vq)=-\psi_{\Gamma,-}(\vq)=\psi(\vq)$ and the
proper Fourier transform of the gap equation keeping gauge
invariance, we find the following equation determining the upper
critical field
\begin{widetext}$$
 \Psi(\vR)\frac{\ln t}{t}=\int_0^{\infty}\frac{du}{\sinh(tu)}\Big \langle
|f_{\Gamma}(\vsk)|^2[\cos(\tilde{\mu}\tilde{\vH}\cdot \hvg
u)+i\cos\delta\sin( \tilde{\mu}\tilde{\vH}\cdot \hvg
u)]e^{-|\hat{v}_{\perp}|^2\tilde{H}u^2/2}e^{\hat{v}_+\Pi_+\sqrt{\tilde{H}}u}e^{-\hat{v}_-\Pi_-\sqrt{\tilde{H}}u}-1\Big
\rangle \Psi(\vR)$$
\end{widetext}
where $t=T/T_c$, $\tilde{H}=Hv_F^2/(2\pi\Phi_0T_c^2)$,
$\tilde{\mu}=\mu_B 2 T_c \Phi_0/v_F^2$, $v_F^2=\langle
v_{\perp}^2\rangle=\langle v_1^2+v_2^2\rangle$, $v\pm=(v_1\pm
iv_2)/\sqrt{2}$, $v_{1,2}$ are components of the Fermi velocity
perpendicular to the magnetic field, $\Pi^{\pm}=(D_1\mp i
D_2)l_H/\sqrt{2}$, $l_H^2=\Phi_0/(2\pi H)$, and $D_i=-i\partial_i-
2e A_i / c$.
The upper critical field $H_{c2}$ is found by expanding
$\Psi(\vR)=\sum_n a_n\phi_n(\vR)$ ($\phi_n(\vR)$ are the usual
Landau levels).

In the following we take a spherical Fermi surface. For CePt$_3$Si
this will not be the case, but the overall geometry of the Fermi
surface does not qualitatively change our results. We also take
$\vg_{\vsk}=\sqrt{3/2}(-k_y,k_x,0)$ as the lowest order term in
$k$ allowed by symmetry and consider the case $V(\vk,\vk')=V_0$
for isotropic $s$-wave pairing. Our results will hold for any
pairing symmetry as the Ginzburg Landau (GL) theory discussed
later will demonstrate.

For the field along the c-axis, $\hvg\cdot\vH=0$ so $H_{c2}$ is
independent of the Zeeman field and there is {\it no paramagnetic
effect} (note that in principle there can be a paramagnetic effect
since there are $\vg_{\vsk}$ allowed by symmetry that contain a
$g_z$ component \cite{sam04-2}, however such terms are expected to
be small). The solution of the upper critical field problem is
identical to that carried out by Helfand and Werthamer
\cite{hel64} and is plotted in Fig.~1. However, for fields
perpendicular to the $c$-axis unusual properties occur, which can
be best illustrated by a GL theory with free energy density
\begin{widetext}
 \begin{equation}
f=- a |\Psi|^2 +\frac{\beta}{2}|\Psi|^4+ \frac{1}{2m}|\vD\Psi|^2 +
\frac{1}{2m_c}|D_z\Psi|^2 + \epsilon \vn\cdot \vB \times [\Psi
(\vD\Psi)^* +\Psi^*(\vD\Psi)]+\frac{\vB^2}{8\pi} \label{eq1}
\end{equation}
\end{widetext}
where $ a=a_0(T_c^0-T)$, $\vD=(D_x,D_y)$, $\vn$ is the unit vector
oriented along the c-axis, and $\vB=\vnabla\times\vA$.
Eq.~\ref{eq1} applies to all possible pairing symmetries with a
single complex order parameter, as discussed also by Samokhin
\cite{sam04-2}. The lack of inversion symmetry allows for the
existence of the term proportional to $\epsilon$ (for a discussion
of other related terms see Ref.~\onlinecite{min94}).  This term
induces a spatially modulated solution in a uniform magnetic
field. The GL equation for the order parameter
$\tilde{\Psi}(\vR)=e^{i\vq\cdot\vR}\Psi(\vR)$  in a field is
identical to the zero-$\epsilon$ GL equation with $\Psi(\vR)$,
where
\begin{equation}
\vq=-2m\epsilon~\vn\times\vB \label{eq3}
\end{equation}
(a microscopic expressions for $\vq$ is given below).
Consequently, the upper critical field solution in the GL limit is
$\Psi(\vR)=\phi_0(\vR)e^{i\vq\cdot\vR}$. We call this phase the
{\it helical vortex phase}. The helical order coincides with an
increase in the upper critical field ($\vB=\vH$):
\begin{equation} T_c(\vH)=T_c-\frac{\pi H}{\Phi_0
\sqrt{mm_c} a_0}+\frac{m\epsilon^2(\vn\times\vH)^2}{2 a_0}.
\end{equation}
The expression for the supercurrent density is
\begin{equation}
\vJ=c\nabla\times\left[\vB-4\pi \vm\right]/4\pi=\vJ_0+4\epsilon
e(\vn\times\vB)|\Psi|^2\label{current}
\end{equation}
where $\vJ_0$ is the usual supercurrent density in the GL limit
and $\vm={\epsilon m\over e}\vn\times \vJ_0$ is the magnetization
due to the $\epsilon$ term Eq.~\ref{eq1}. The usual boundary
condition on the order parameter is given by $\vJ=0$ through the
boundary. The appearance of $\vm$ in Eq.~\ref{current} is highly
unusual in GL theory and has some consequences that are discussed
later. Note that if $|\Psi|$ and $\vB$ are spatially uniform then
$\vJ=0$ for $\vq$ given by Eq.~\ref{eq3}; the helical phase
\emph{carries no current}. The possibility of this helical order
has been raised in the context of thin film or interface
superconductivity where the vector potential can be neglected
\cite{ede89,agt03,dim03}. As discussed below, we find here that it
can play a very important role in the vortex phase.

\begin{figure}
\epsfxsize=2.5 in \center{\epsfbox{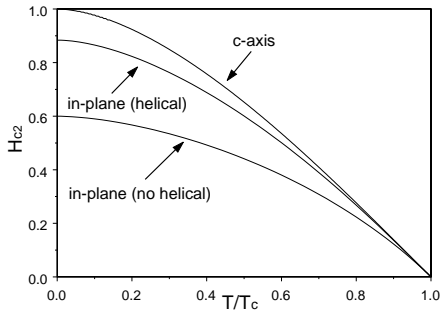}} \caption{Upper
critical fields for CePt$_3$Si with fields along the c-axis and in
the plane. The actual in-plane $H_{c2}$ will lie between the two
extremes shown. The effect of the helical order on $H_{c2}$ can be
quite pronounced. These calculations are for $\alpha_M=3$.}
\label{fig1}
\end{figure}

The increase in $H_{c2}$ due to the appearance of the helical
plays an important role in the microscopic theory. Since $\vq
\cdot \vH =0 $, we can expand $\phi_0(\vR)e^{i\vq\cdot\vR}=\sum_n
b_n(q)\phi_n(\vR)$ where
$b_n=(iql_H)^ne^{-(ql_H)^2/4}/\sqrt{2^nn!}$. Our numerical
microscopic  solution has this form and near $T_c$ we find,
\begin{equation}
q=2\mu_BH\cos\delta\frac{\langle \hvg(\vk)\cdot\mhx
v_{F,y}(\vk)|f_{\Gamma}(\vk)|^2\rangle}{\langle
v_{F,y}^2|f_{\Gamma}(\vk)|^2\rangle}\label{q},
\end{equation}
for $f_{\Gamma}(\vk)=1$ this gives
$ql_H=0.373\alpha_M\cos\delta\sqrt{H/H_{c2}^0}$ where
$\alpha_M=\sqrt{2}H_{c2}^0/H_P$ is the Maki parameter, $H_{c2}^0$
is the upper critical field for $\mu_B=0$ (this coincides with the
$H_{c2}$ for the field along the $c$-axis), and
$H_P=\Delta/(\mu_B\sqrt{2})$. For $f_{\Gamma}(\vk)=k_x^2-k_y^2$,
Eq.~\ref{q} gives $ql_H=0.418\alpha_M\cos\delta\sqrt{H/H_{c2}^0}$.
The enhancement of $H_{c2}$ due to the helical order can be
substantial as Fig.~1 shows. Our numerical results show $ql_H$ can
be larger than one, which implies that the helical wavelength
becomes less than the spacing between vortices. Fig.~1 is for
$\alpha_M=3$ which is slightly smaller than $\alpha_M=3.8$ that
follows from the measurements of Ref.~\onlinecite{yas04}. The
helical order changes with varying $\cos\delta$. For
$\cos\delta=0$ the density of states are the same on both Fermi
surfaces and no helical order appears (labelled `in-plane (no
helical)' in Fig.1); while for $\cos\delta=1$, $ql_H$ is maximum
(this corresponds to the curve labelled `in-plane (helical)' in
Fig.~1). For all other possible values of $\cos\delta$, the
$H_{c2}$ curve lies between these two extremes. The limit
$\cos\delta=1$ is unlikely since this implies that the density of
states of one of the two bands vanishes. However, if the elements
of the pairing interaction $V_{\alpha,\beta}$ in
Eq.~\ref{twoband2} are different in magnitude from each other,
then a large $ql_H$ can still arise for $\cos\delta=0$.

Comparing our result with the $H_{c2}$-measurement by Yasuda {\it
et al.} \cite{yas04} for CePt$_3$Si, we find that only taking the
effect of the spin-orbit coupling  into account would not account
for the relatively large value of the in-plane $H_{c2}\approx 2.8
T $) at $T=0$. Here, paramagnetic limiting should reduce the value
to below 2 T. We can, however, explain the increased $ H_{c2} $ by
including the helicity.


We have also examined the behavior of the Abrikosov parameter
$\beta_A=\langle|\Psi(\vR)|^4\rangle/\langle|\Psi(\vR)|^2\rangle^2$
in connection with the possible vortex lattice structures. We find
a possible structural transition from a stretched hexagonal
lattice at high temperatures to a stretched square lattice at low
temperatures. The origin of this transition is related to the
two-dimensional inhomogeneous state discussed in
Refs.~\onlinecite{bar02,dim03,ada03}. Note that the helical order
discussed above is distinct from that discussed in these works.
The physics discussed in Refs.~\onlinecite{bar02,dim03} does not
play a significant role in CePt$_3$Si because the value of
$\alpha_M$ is too small. However, if $\alpha_M>>1$ then the vortex
physics becomes very exotic.


We have not discussed the direct experimental verification of the
helical phase. Since helicity of the order parameter is related to
its phase, an interference experiment based on the Josephson
effect would provide the most reliable test. Here we propose to
consider a Josephson junction between two thin film
superconductors (Fig.~\ref{fig1}), one (1) with and the other (2)
without inversion symmetry (for CePt$_3$Si the c-axis is
perpendicular to the film). For a magnetic field applied in the
plane of the film {\it perpendicular} to the junction and with the
superconductor (2) is oriented so that the helicity $ \vq $ is
perpendicular to the field ; we find this gives rise to an
interference effect analogous to the standard Fraunhofer pattern.
For this experiment, the film must be sufficiently thin that the
magnetic field and the magnitude of the order parameter are
spatially uniform.

To illustrate this, consider the following free energy of the
junction
\begin{equation}
H_J=-t\int dx[\Psi_1(\vR)\Psi_2^*(\vR)+c.c.]
\end{equation}
where the integral is along the junction. The resulting Josephson
current is
\begin{equation}
I_J=Im\Big [ t\int dx\Psi_1(\vR)\Psi_2^*(\vR)\Big ]
\end{equation}
Setting the junction length equal to $2L$, and integrating yields
a maximum Josephson current of
\begin{equation}
I_J=2t|\Psi_{1}^0||\Psi_{2}^0|\frac{|\sin(qL)|}{|q L|}
\end{equation}
This, combined with the result of the microscopic theory near
$T_c$ (for an isotropic interaction) that
$qL=2.4\cos\delta\alpha_MH\xi_0L/\Phi_0$ (where
$\xi_0=0.18v_F/T_c$) demonstrates that the Josephson current will
display an interference pattern for a field {\it perpendicular} to
the junction. Note that in the usual case the Fraunhofer pattern
would be observed for a magnetic field perpendicular to the thin
film. Furthermore, for a sufficiently large in-plane field, so
that $|qL|>>1$ (which implies $I_J\approx0$), the arguments of
Ref. \onlinecite{yang00} imply that the application of an
additional magnetic field along the surface normal will lead to an
asymmetric Fraunhofer pattern.

\begin{figure}
\epsfxsize=3.5 in \center{\epsfbox{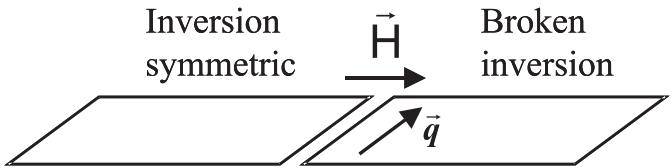}} \caption{Josephson
junction geometry for the observation of a helical phase.}
\label{fig1}
\end{figure}

A less direct probe of the helical order is to look for the
related transverse magnetization that appears in
Eq.~\ref{current}. There are two situations for which this can be
observed. The first is in a thin film with a supercurrent flowing
along $\hat{x}$ and a surface normal along $\hat{y}$. In this case
a magnetization will exist along $\hat{y}$ (normal to the film).
This situation is a generalization of an experiment originally
proposed by Edelstein \cite{ede95}. We estimate $|\vm|={3\pi\over
4}\cos\delta n_s \mu_B v_s/v_F\sim 0.02$ Gauss assuming
$v_s/v_F=2\times 10^{-4}$, $\cos\delta=1/3$, and $n_s=1\times
10^{28}$ m$^{-3}$. The second is in the vortex lattice phase for a
field applied along the c-axis. In this case, a calculation valid
near $H_{c2}$ gives $\vm={\epsilon m\over e}\vn\times \vJ_0$,
where $\vJ_0$ is the usual supercurrent for the Abrikosov vortex
lattice. Near the vortex core, $\vm$ is directed radially outward,
perpendicular to the applied field.

We have considered the role of magnetic fields on the
non-centrosymmetric superconductor CePt$_3$Si. Using a two-band
theory with a Rashba spin-orbit interaction, we have shown that
the upper critical field for the field along the c-axis behaves as
if it would in a conventional superconductor {\it independent} of
the paramagnetic (Zeeman) field. We have further shown that while
there is a paramagnetic limiting effect for magnetic fields
applied in the basal plane, this effect can be strongly reduced by
the appearance of a helical order. Our theory agrees with the
experimental measurements of $H_{c2}$, despite a relatively strong
Zeeman field, provided this helical order exists. Finally, we have
proposed a Josephson junction experiment that can unambiguously
identify the helical order and discussed the appearance of a
transverse magnetization related to the helical order.

We are grateful to E. Bauer, P. Frigeri, A. Koga, T.M. Rice, and
K.V. Samokhin for many helpful discussions. This work was
supported by the Swiss National Science Foundation. DFA and RPK
were also supported by the National Science Foundation grant No.
DMR-0381665, the Research Corporation, and the American Chemical
Society Petroleum Research Fund.

\end{document}